\documentclass[conference]{IEEEtran}
\usepackage[latin1]{inputenc}

\usepackage{epsf}\epsfverbosetrue

\usepackage{graphics,epsfig,color}

\usepackage{graphicx,epsfig}

\usepackage{multirow}

\usepackage{slashbox}

\usepackage{alltt}

\usepackage{subfigure}

\usepackage{color}

\usepackage{float}

\usepackage{url}

\usepackage{verbatim} 

\usepackage{footnote} 

\usepackage{latexsym} 

\usepackage{amsmath}

\usepackage{amsfonts}

\usepackage{amssymb}

\usepackage{sidecap} 

\usepackage{color}

\usepackage{wrapfig}
\usepackage{changebar}

\ifCLASSINFOpdf
\else
\fi
\begin{document}
%
\title{NetInf Mobile Node Architecture and Mobility Management based on LISP Mobile Node}


\author{{Muhammad Shoaib Saleem, \'Eric Renault \& Djamal Zeghlache}\\
\IEEEauthorblockA{Institut T\'el\'ecom \ - T\'el\'ecom SudParis, \'Evry, France\\
Email: \{shoaib.saleem, eric.renault, djamal.zeghlache\}@it-sudparis.eu}
}


%


\maketitle

\begin{abstract}
In this paper, we propose an architecture for Network of Information mobile node (NetInf MN). It bears characteristics and features of basic NetInf node architecture with features introduced in the LISP MN architecture. We also introduce a virtual node layer for mobility management in the Network of Information.                                                                                                                                                                                                                                                                                                                                                                                                                                                                                                                                                                                                                                 Therefore, by adopting this architecture no major changes in the contemporary network topologies is required. Thus, making our approach more practical.
\end{abstract}


%
\IEEEpeerreviewmaketitle

\section{Introduction}

4WARD~\cite{4ward} is the proposed architecture and design for the Future Internet. The Network of information (NetInf) is one of the components of the 4WARD project. The cornerstone of this architecture is that the information takes the prime position superseding the node-centric approach of the current internet. The genesis of NetInf was influenced by existing major technologies and beyond. The features which NetInf exhibits are the melange of existing and innovative solutions. 

Mobility is one of the scenarios considered in NetInf. Late Locator Construction (LLC)~\cite{LLC} is a proposal to handle mobility and multihoming issues in NetInf. It implements the locator/identifier separation idea. Whenever there is a mobility or rehoming event, there is no interruption in the ongoing TCP session between nodes. The notion behind the LLC is to use a Global Locator (GL) for routing between the Core Network (CN) and the Edge Networks (ENs). IPv4/IPv6 can be considered as the CN and nodes, mobile or stationary, forming a topology at the edge of CN, can be considered as ENs. GL is built inside the Locator Construction System (LCS) which is embedded inside the core network. Each Edge Network (EN) and the host attached to these ENs have Attachment Registers (AR) within the LCS. Each AR has IDs of itself and of its neighbors. The overall goal of LLC is to minimize the update signaling to the Locator Construction System (LCS) to update the GL to deal with the scalability problem within the core network. It has also been proposed that the core network scales well as core and edge networks are responsible for their own routing~\cite{LLC}.

Location Identifier Separation Protocol (LISP)~\cite{lisp} is a network based approach for locator/identifier separation. It focuses on limiting the size of routing tables and improving scalability and routing system. LISP Mobile Node (MN)~\cite{lispmn} is an extention to the classic LISP for mobile nodes. It has multiple design goals including a wide range of communication possibilities in different mobility cases along with multihoming in MN, as well as allowing MN to act as a server. The LISP MN architecture has some LISP features together with additional characteristics to support mobility and rehoming events. What has been envisaged here for NetInf is to provide seamless connectivity between the mobile nodes even if there is a simultaneous roaming. Embedding LISP MN features in NetInf MN can then improve mobility management in LLC for the NetInf architecture.


 

\section{Mobility Management in NetInf}

Although LLC is a good proposal, it has however some discrapencies in terms of properly addressing both scalability and mobility issues. LISP MN inherits classic LISP features but lacks complete compatibility to work with LLC. Our proposal consists in a new approach to deal with these issues. The prime goal is the mobility management in NetInf. A NetInf MN should bear charactersitics and features of classic NetInf node architecture together with features introduced by the LISP MN architecture. The design goal of such a node is to deal with mobility in a highly dynamic environment.

 \subsection{NetInf MN Architecture}

Fig. 1 presents a high-level overview of the NetInf MN architecture. At the bottom both physical and network layers provide services to the transport layer placed above. Within the transport layer is located the transport control engine (TCE). In NetInf, TCE is responsible for the coordination of protocols used for accessing NetInf objects. In this design, both Inner Locator Construction Tunnel Router (ILCTR) and Outer Locator Construction Tunnel Router (OLCTR) include their functionalities in TCE. The goal of these two routers is to work under the conditions when non-NetInf sites are communicating.
\begin{figure}[hb]
\begin{center}
{
\includegraphics[width=6cm, height=4cm]{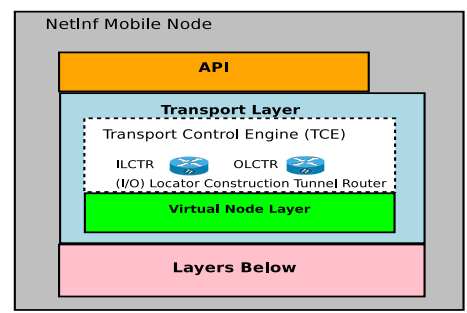}
}
\caption{NetInf Mobile Node Architecture.}
\label{fig_1}
\end{center}
\end{figure}
\vspace{-0.5cm}
 The working of these routers will be extensively explained in the future work. The virtual node layer within the TCE is another feature introduced for the mobility management of information. At the top, the API provides services to users. This API gets services from the transport layer below. It should be noted that NetInf and the nodes are interdependent on each other. Nodes are dependent on NetInf for the network management, while NetInf depends on nodes in scenarios where it has to manage and propagate the resources inside the network.

\vspace{-0.1cm}


\subsection{Mobility Scenarios}

In order to understand the basic functionalities of our proposed architecture, consider two mobility scenarios. In Fig. 2, MN1 roams into a new edge network EN3 during the on-going session with a stationary node MN2 in edge network EN2. MN1 first contacts local server LS in EN3 to register its new ID allocated by the DHCP. It also queries the local cache mechanism to find any trace of MN2 activities in EN3. If it is successful in finding one, it can use that cached locator and resume its session. Otherwise, it contacts the core network. 










Fig. 3 presents a simultaneous rehoming event. Two mobile nodes MN1 and MN2 are communicating in their respective edge networks EN1 and EN2. A mobility event occurs when both mobile nodes move to a new edge network EN3 at the same time. In this case, it would be more efficient if mobile nodes MN1 and MN2 could access a local server in EN3 and register their IDs to continue their session, instead of connecting to the core network. As IDs are global and unique, no further name resoluton is then required. However, a lookup request must be sent to the local server when both nodes are registered and each node should set TTL. If nodes relocate each other before their TTL expires, the session is re-established (in case of interruption) or continued. Otherwise, both should contact the core network.

\section{Mobility Management of Information in NetInf through Virtual Node Coordination}
The main problems faced in the network because of mobility are: the unpredictable motion of nodes and unpredictable availability of nodes i.e. nodes are continuously joining and leaving the network. Since information takes the centric position within the network, the mobility management of the information becomes a major concern. One solution for the management of information mobility is the use of virtual mobile nodes. These virtual nodes can be thought of as a program running on real nodes. Whenever there is a mobility or rehoming event, the node which departs from the local network handovers its connectivity to one of the virtual nodes so that the connectivity of the ongoing session is carried away smoothly. This is made possible in our proposed NetInf MN architecture by the introduction of a new layer, the Virtual Node layer that provides virtual mobile node services in coordination with the real node. All mobile nodes can act as temporary virtual mobile node and can provide services to the local mobile nodes.

For example, in Fig. 2 EN1 hosts two nodes namely MN1 and MN3. These two nodes have the ability to support virtual mobile node features introduced in their respective layered architecture. Lets assume that MN1 is in session with MN2 in EN2. When MN1 moves from EN1 to EN3, it initiates the algorithm to make use of the embedded virtual layer. It contacts MN3 in EN1 and establishes a local connection by informing about its new destination (EN3). MN3 is now acting as a virtual MN1 and is communicating with MN2. This continues until MN1 reaches EN3. When this happens, it gets all the updates from MN3 and reconnects to MN2 again.


\vspace{-0.1cm}
\begin{figure}[hbtp]
\begin{center}

{

\includegraphics[width=9cm, height=4cm]{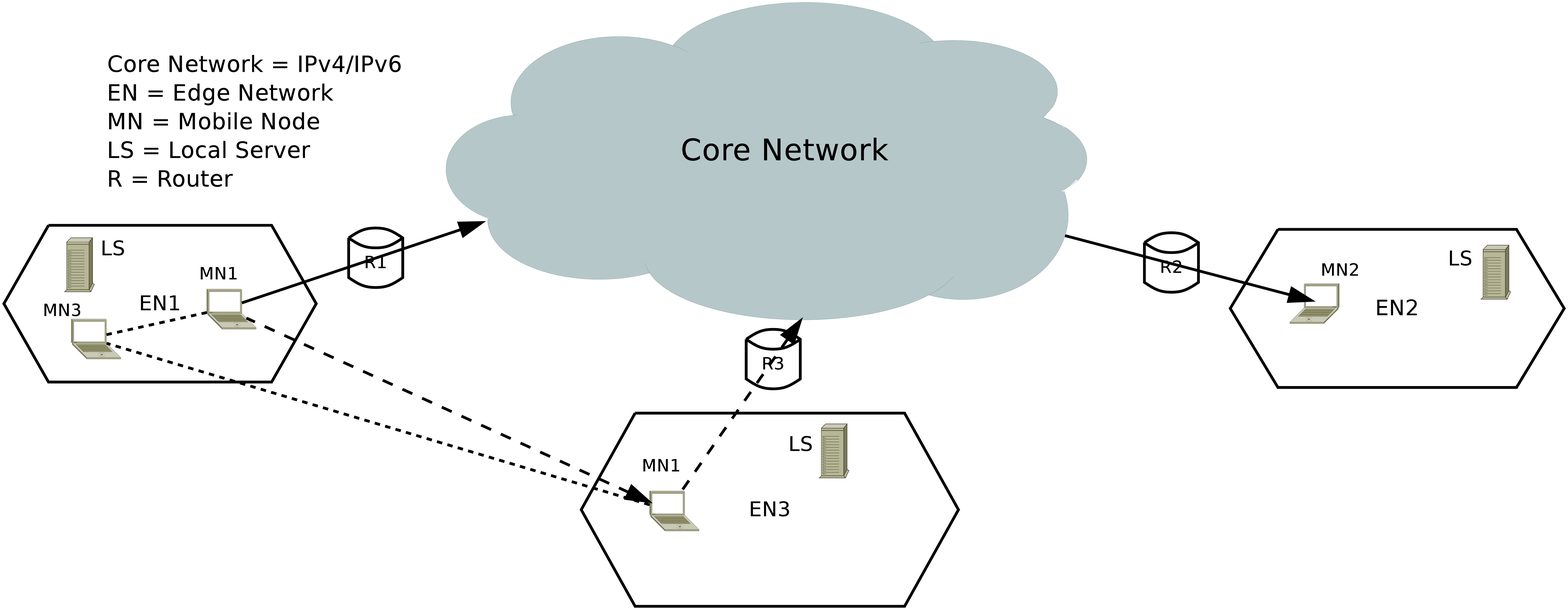}

}

\caption{Single Mobile Node Roaming.}

\label{fig_1}

\end{center}

\end{figure}

\vspace{-0.8cm}
\begin{figure}[hbtp]

\begin{center}

{

\includegraphics[width=9cm, height=4cm]{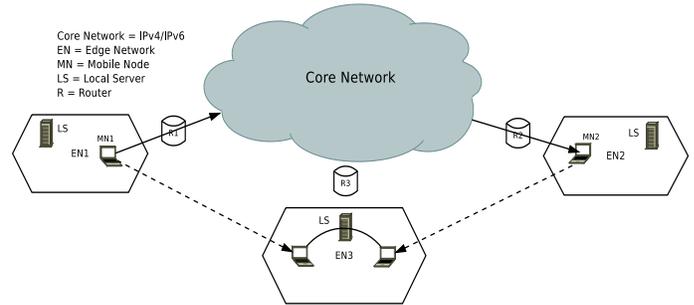}

}

\caption{Simultaneous Mobile Nodes Roaming towards same Edge Network.}

\label{fig_1}

\end{center}

\end{figure}
\vspace{-0.6cm}
\section{Conclusion and Future Work}


There are still some open issues to be addressed for the proposed architecture for mobile entities in  NetInf. So far, the work is on the midway. ILCTR and OLCTR services are yet to be finalised. Besides providing services for non-NetInf sites, they can also work as storage spaces in edge networks in case of disconnection from the core network. Different parameters have to be considered to evaluate the proposal. As ILCTR encapsulates packets before sending them to non-NetInf sites, how could the Maximum Transmission Unit (MTU) be maximized. A new efficient protocol shall be proposed to deal with the mobility cases presented above. At present, we are in the process of evaluating our architecture through extensive simulations and real mobility traces. Finally, we plan to address the multihoming issue along with the security issue during mobility.

\begin{thebibliography}{}


\bibitem{4ward} \textit{The FP7 4WARD Project}. http://www.4ward-project.eu/

\bibitem{LLC} Eriksson, A. and Ohlman, B, \emph{Dynamic Internetworking Based on Late Locator Construction},
IEEE Global Internet Symposium, 2007.

\bibitem{lisp} D. Farinacci, V. Fuller, D. Meyer, and D. Lewis, \emph{``Locator/ID Separation Protocol (LISP)''},
http://tools.ietf.org/html/draft-ietf-lisp-06, Jan. 2010.

\bibitem{lispmn} D. Farinacci, V. Fuller, D. Lewis, and D. Meyer , \emph{``LISP Mobile Node''},
http://tools.ietf.org/html/draft-meyer-lisp-mn-01, Feb. 2010.




\end{thebibliography}
\end{document}